\documentclass[twocolumn]{aa}
\usepackage{graphicx}
\usepackage{txfonts}
\begin{document}
 
\title{Systematic effects induced on IBIS detectors by background and inhomogeneity 
 of the spatial response 
  \thanks
     {Based on observations with INTEGRAL, an ESA project with instruments
     and science data centre funded by ESA member states (especially the PI
     countries: Denmark, France, Germany, Italy, Switzerland, Spain), Czech
     Republic and Poland, and with the participation of Russia and the USA.}
 }

\titlerunning{IBIS background and non-uniformity effects}

\author{L. Natalucci\inst{1}, A.J.Bird\inst{2}, A.Bazzano\inst{1}, P.Ubertini,\inst{1},
        J.B.Stephen\inst{3}, R.Terrier\inst{4}, L.Lerusse\inst{5}  
          }
\authorrunning{Natalucci et al.}

\offprints{L. Natalucci}

\institute{CNR-Istituto di Astrofisica Spaziale e Fisica Cosmica, 
           Area Ricerca Roma 2/Tor Vergata, Via del Fosso del Cavaliere 100,
           00133 Roma, Italy\\
              \email{lorenzo@rm.iasf.cnr.it}
         \and
             Department of Physics \& Astronomy, University of Southampton, SO17~1BJ,U.K.\\ 
	 \and 
	   CNR-Istituto di Astrofisica Spaziale e Fisica Cosmica,
	             Area Ricerca di Bologna, Via Gobetti 101, 40100 Bologna, Italy \\
	 \and
	   DAPNIA, Service d'Astrophysique, CEA/Saclay, 91191 Gif-sur-Yvette Cedex, France\\
	 \and
	 INTEGRAL Science Data Centre, Chemin d'Ecogia 16, CH-1290 Versoix, Switzerland \\ 
	     }

\date{Received; Accepted}

\abstract{
The spatial distribution of the background events may
affect the source detection capability of IBIS at high
energies ($\geq200$~keV) for both ISGRI and PICsIT layers.
The observed background is found to be variable
and spatially structured, and in some cases its properties
strongly deviate from the expected statistical behaviour.
Background correction methods are then necessary to
improve the quality of the shadowgrams obtained from sources.
In order to perform an efficient flat-fielding the
response of the detector to both source ($\gamma$-rays) and
background events is investigated using data from Monte 
Carlo simulations and in-flight calibration observations.

         }
\maketitle

%

\section{Introduction}

The IBIS instrument on board {\em INTEGRAL} (\cite{winkl}) is the first
large area, space born {$\gamma$}-ray telescope carrying a pixellated, 
multi-layer detector system (\cite{huber}). Due to its large 
aperture and large collecting area, the scientific 
performances are strongly dependent on background reduction. 
Optimal levels of background intensity and 
stability in time have been reached after tuning and 
calibration of instrumental parameters. This has been achieved 
during in-flight commissioning (\cite{huber}). 
In IBIS, different data acquisition modes working simultaneously 
ensure a good capability of background rejection. Important features 
are the selection of single and multiple events in the high energy 
detector (PICsIT, \cite{ofcok}) and of multi-layer coincidence events 
({\em Compton} mode). Furthermore, the two detector layers 
ISGRI (the hard X-ray detector, \cite{lebru}) and PICsIT are actively
shielded by a multi-module VETO system, based on one lateral and 
one bottom arrays of BGO blocks (\cite{quadr}). ISGRI is also shielded 
from the diffuse hard X-ray background outside the field-of-view by a 
composite passive shield protecting the whole telescope from mask 
to detector base (see e.g. \cite{natca}). 

The IBIS data processing software (\cite{gold1}) is
capable of recovering many time varying systematic effects as 
telemetry losses,
noisy pixels, temporary switch-off of ISGRI modules, 
dead time etc., which are well quantified and properly taken 
into account in the analysis. Once the data have been corrected 
for  these effects we still see important spatial structures, which 
affect both detector layers. These are easily detected as  
different average count-rates in 
detector modules/semi-modules, enhancements at the edge of the modules 
and local effects induced by the readout electronics. 
There are several factors known to produce this inhomogeneity of the
spatial response (see also \cite{steph}; \cite{terri}): 
a) a spatially dependent hadronic background component, depending on 
the external payload and spacecraft mass distribution; 
b) the efficiency of detection of single and multiple events, 
which is expected to be dependent on the detector position; 
c) the individual response of each detector module; d) the intrinsic
response of each pixel as a function of its position, i.e. its 
proximity to a module, an ASIC (Application Specific Integrated 
Circuit) readout element and/or a dead pixel; e) variations 
of the performance of the VETO modules, induced especially by 
temperature variations. 
These effects are, in turn, energy dependent. Moreover a given pixel 
may undergo periods of unusual behaviour. For ISGRI, defective pixels 
including noisy elements which are switched off permanently or for a 
relatively short time period may induce local variations in the detector 
itself and even in PICsIT.
PICsIT pixels have also intrinsic noise levels that could have local 
effects in ISGRI. PICsIT data are normally transmitted as 
{\em spectral histograms} (SI) already accumulated on board. 
Varying count rates of pixels can influence the local response of
detectors for less than the typical SI integration time ($\approx2150s$),
and hence cannot be properly followed in PICsIT. For this reason, local
corrections based on analysis of single pixel light curves cannot be 
always applied efficiently. Therefore, methods of flattening the images
independently of the time behaviour of pixels are being studied
(see Sect.~2).

The difficulty of correcting data by using maps obtained by long 
background exposures 
is mainly related to the background being variable with time, and to 
different systematics in the counting behaviour of pixels in 
"observation" and "background" exposures. In the following sections,
an overview of the main problems related to background induced effects
is given and some results are reported in more detail, especially  
regarding the PICsIT detector.
For PICsIT, an important source of background is the excess
counts caused by the decay of long-lived phosphorescence
states of the CsI crystals (\cite{hurle}).  
These counts cannot be vetoed because they give 
rise to single events, separated by intervals of the order of a 
few~$10^{-5}$s. However, they can be easily identified and removed 
in {\em photon by photon} data, and their spatial distribution can 
be separately studied (see Sect. 4). Other spatial inhomogeneity 
effects which are background independent are described in Sect. 3.

\section{Flat-fielding methods}

\begin{figure}
\centering
\includegraphics[width=8.5cm]{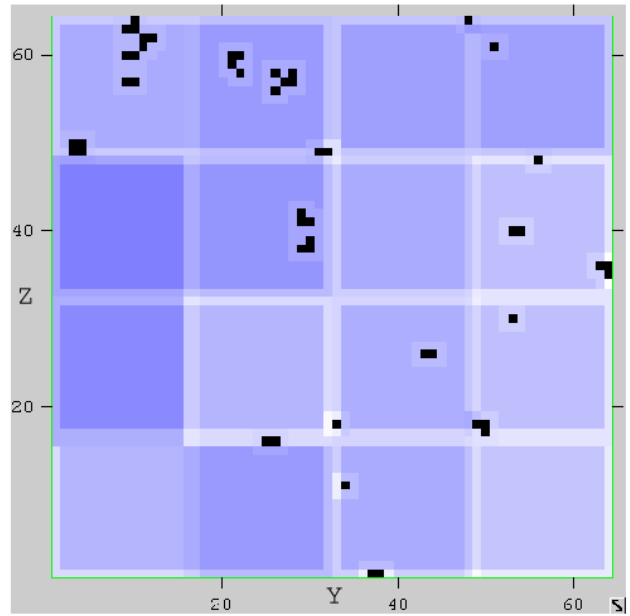}
\caption{Example of PICsIT model map in the energy range 
          190-300 keV. It shows different average levels associated
	  to the 16 semi-modules, the {\em edge} pixels, and pixels
	  surrounding {\em dead} pixels (the latter are shown in 
	  black). The image axes (in units of pixel index) are parallel 
	  to the spacecraft
          Y- and Z-axes, for which the X-axis represents the
	  pointing direction.
}
\label{Fig1}
\end{figure}

For IBIS, flat-fielding prior to image deconvolution (\cite{goldw})
is essential especially at high energies,
as long as the size of the fluctuations in the background level, 
propagated through the imaging process exceed the observed celestial source 
intensity. The approach of applying a flat-field based on Fourier filtering 
is difficult, because a substantial power of the background distribution 
is seen at high spatial frequencies, where most of the signal from 
multiplexing the source counts by the coded mask is also found. 
An alternative approach is 
to characterize each pixel by averaging on pixel categories, defined on the 
basis of the pixel position in the detector, i.e. proximity to a module edge 
and/or to a noisy/dead pixel, position within an ASIC etc. This
has the form of a model map applicable to a given energy band. The model 
is obtained from real data by measuring average count rates for each pixel 
category (see Fig.~\ref{Fig1}).
Correction maps defined for eight standard 
energy bands, covering the range 0.2-6.5 MeV, have been used
to subtract a background shape from the single event shadowgrams.
The resulting detector image is defined by a subtraction of the model from 
the original image: 
${D}_{ff}$=(${D}_{raw}$-$Mf$)+$\langle {D}_{raw} \rangle$, where $M$ is
the model map and $f$=$\langle {D}_{raw} \rangle$/$\langle M \rangle$
is the ratio between average 
pixel values in the raw image and model map. 
								       

The standard deviation in the pixel counts of the background subtracted 
shadowgram is compared to that of the original image, and a {\em flat-field} 
efficiency is evaluated as: 
\vspace{0.2cm}

${\epsilon}_{ff}$~=~(${\sigma}_{raw}-{\sigma}_{ff}$) / (${\sigma}_{raw}-\sqrt{\langle {D}_{raw} \rangle}$)

\begin{figure}
\centering
\includegraphics[width=8.5cm, height=6.9cm]{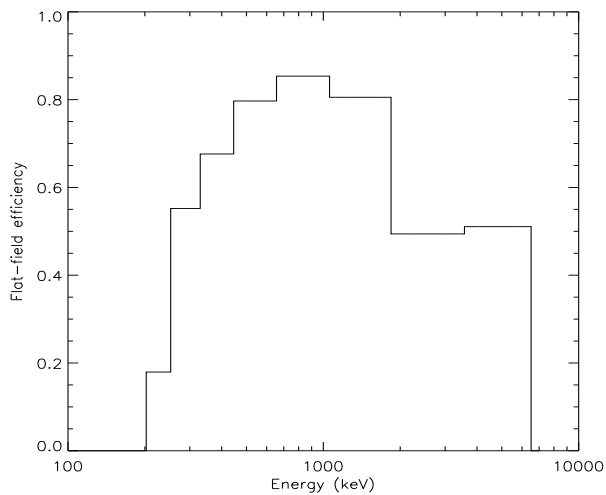}
\caption{Flat-field efficiency as a function of energy, for 
         the model maps discussed in Sect.2. 
        }
\label{Fig2}
\end{figure}

\vspace{0.2cm}
We have used a model map based on the sum of 3 science windows, for a
total exposure of 6450 s of an empty
field observation, to flatten   
a series of spectral histograms obtained from a Crab exposure.
In Fig.~\ref{Fig2}, the average flat-field efficiency obtained
by this processing is plotted
as a function of energy. We remark that in the first two energy 
bands, the low efficiency obtained is not intrinsic to the flat-field
process, but is rather due to the effect of phosphorescence counts 
(see Sect.~4), causing the average deviates to be well above 
the Poisson statistical value. Apart from this effect, the 
efficiency is energy dependent. 

%
Adding more input data in building the model maps is not found to improve 
the efficiency. This is probably due to the time variations in 
the background. 
Improvements to the model are currently under study, 
as long as more systematic effects are identified.


\section{Response to {$\gamma$}-ray source}

\begin{figure}
\centering
\includegraphics[width=8.0cm,height=6.9cm]{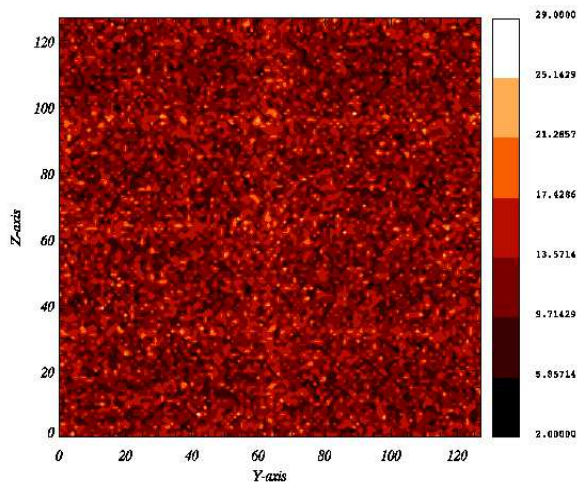}
\caption{Simulated map of ISGRI, as illuminated by a monochromatic source
         at 500 keV. The energy range is 170-400 keV. The image shows a 
	 non perfectly flat events distribution. Both axis scales refer to
	 pixel index (0 to 127 on each side). Axes are parallel to the 
	 spacecraft
	 Y- and Z-axes.
        }
\label{Fig3}
\end{figure}

Besides the known effect of background non-uniformity, scattering of 
celestial source {$\gamma$}-rays associated with a non-ideal response of the 
event selection logic and/or non-ideal geometry of the detector can produce 
a non-homogeneity of the source count distribution. This component of the 
non-uniformity (which is intrinsically different from the background 
induced one) is being studied by Monte Carlo simulations using 
a numerical code based on a detailed geometric and mass model of the telescope,
including electronics and event selection logic (\cite{laure}). 
For this purpose, we have simulated a 
parallel flux of {$\gamma$}-rays incident on the IBIS detector with zero~$\deg$ 
inclination, at different monochromatic photon energies and for a power 
law source. The efficiency and response of single pixels are considered 
uniform in the simulation, so inhomogeneities cannot be ascribed to their
local behaviour. For a monochromatic source, it was verified that the image is 
perfectly uniform when selecting events under the full-energy peak. This
is expected as far as there is no re-distribution of the event energy in
pixels other than the incidence pixel. In contrast, the count distribution
shows differences, 
especially in the counting rate at the edge of modules, 
when the energy range selected covers part of the Compton continuum.
This is true in
particular for ISGRI (see Fig~\ref{Fig3}) and PICsIT multiple events. This 
non-uniformity should be ideally corrected by dividing the residual detector 
image after background subtraction or flat-field (see Sect. 2) by 
an {\em efficiency} map
which takes into account the scattering induced effects. This is, of
course, dependent on the input source spectrum.  A set of reference 
efficiency maps are being computed for a limited set of input 
spectra models (based on power law or cutoff power laws) to be tested 
and compared. We consider, however, that  
correcting real data using a non-uniformity model of the source 
illumination is expected to be of secondary importance compared to  
an efficient background subtraction.

\section{Effects of phosphorescence induced events}

\begin{figure}
\centering
\includegraphics[width=8.2cm]{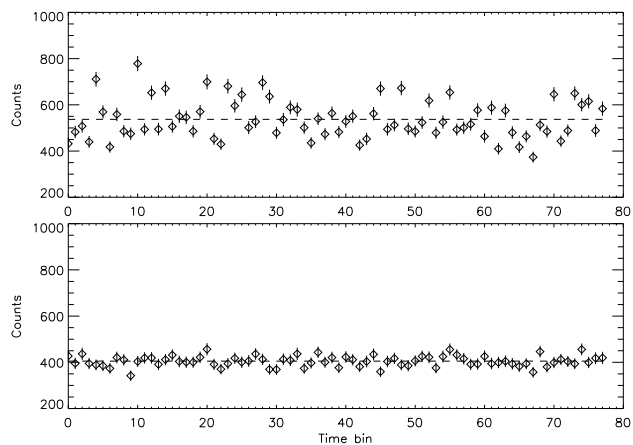}
\vspace{0.5cm}
\caption{Light curves of a sample of {\em photon by photon} data for
         PICsIT single events. Events are rebinned on 100s. 
	 Plotted errors are the expected Poisson 
	 uncertainties (1~$\sigma$). On the upper plot, the total light 
	 curve shows important flickering. This is not present when 
	 phosphorescence events are removed from the data (lower plot). 
        }
\label{Fig4}
\end{figure}

\begin{figure}
\centering
\includegraphics[width=8.5cm,height=10cm]{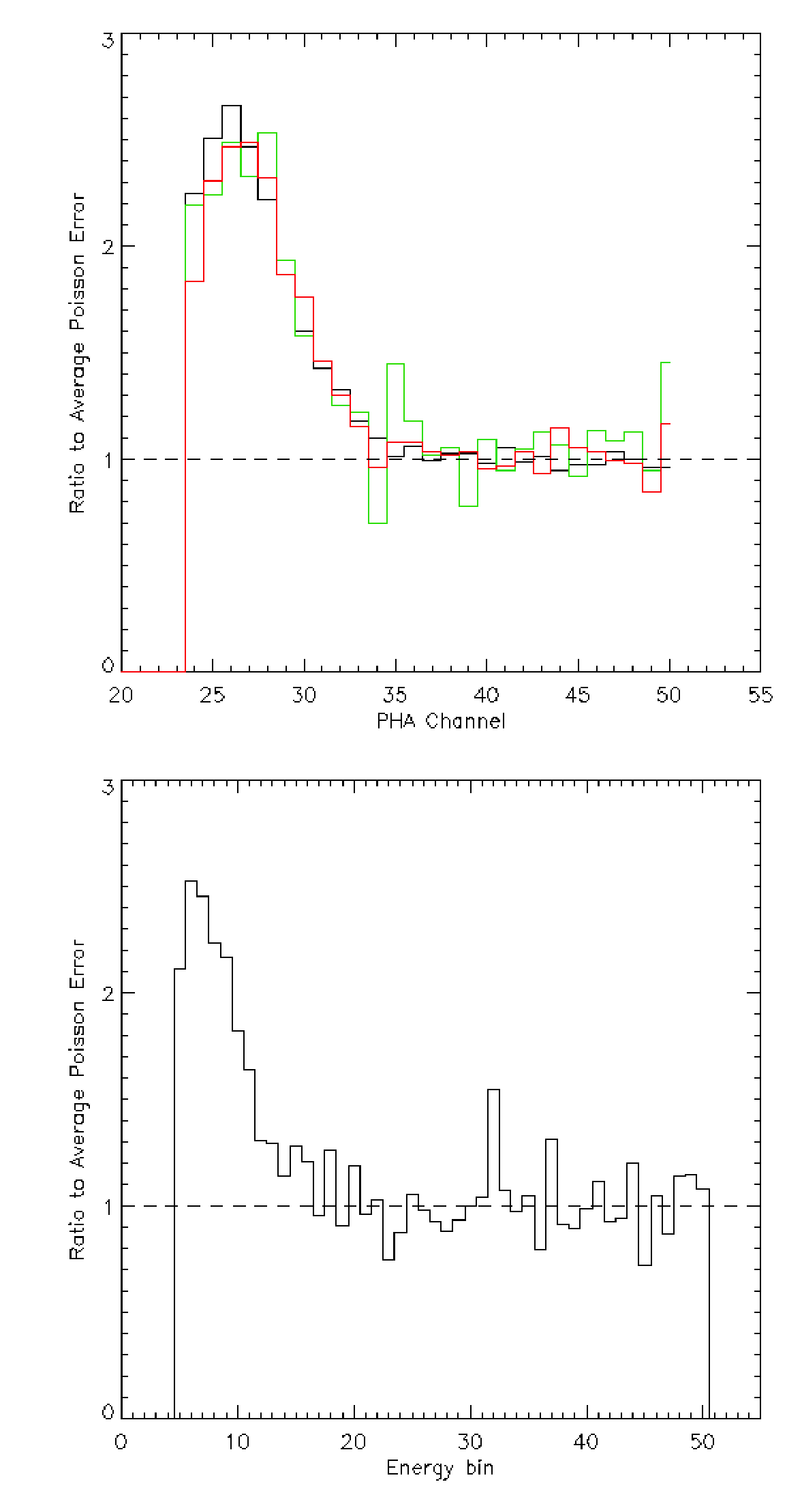}
\vspace{0.3cm}
\caption{ The ratio of the observed standard deviation to the Poissonian expected
value, as a function of energy bin, for PICsIT {\em photon by photon} 
data (top) and for PICsIT {\em spectral histograms} (bottom). The three curves 
in the upper panel refer to different integration times: 10s (black), 
100s (red), 500s (green). The integration time of histograms is 
fixed to 2150s. In the upper panel, data are plotted against PHA channels, 
whilst in the lower panel, SI energy bins are used. Both X-axis scales are linear 
in energy, with a conversion factor $\approx7$~keV/bin. Due to the different
offset, a value of 250~keV corresponds
to abscissa values of 36 and 16 in the two plots, respectively. 
        }
\label{Fig5}
\end{figure}

The PICsIT detector is found to show significant count-rate increase caused 
by {\em fake} events, induced by decay of phosphorescent states in the CsI 
crystals. These states are excited by passage of particles in 
the active body of the PICsIT detector, which occur at a rate $\sim1000$ 
times lower than the {\em normal} background rate. One {\em burst} 
of fake events can be produced by either the passage of a primary particle, 
or by a secondary shower (cascade) producing a track or other figure of 
well-defined spatial pattern (\cite{segre}). All these counts are mostly 
detected at energies below $\sim300$ keV and can be easily recognized in 
{\em photon by photon} data, in which we have full timing information. 
Since each individual passage of particles produces a relatively large number 
of these events, the count rate statistics is strongly modified.
The two panels of Fig.~\ref{Fig4} show the fluctuations of the summed count rate 
for 16 pixels of an ASIC, in the energy range 170-285 keV.  On the upper plot, 
representing the total counts, the fluctuations are clearly non-Poissonian. 
Once the phosphorescence triggers are eliminated, the fluctuations are fairly 
compatible with Poissonian noise (lower panel). It is quite easy to recognize, 
by comparing the two plots, that some time bins are not affected at all
by phosphorescence triggers. This effect can be quantified 
by computing the ratio, $R$, between the observed standard deviation of the 
counts and the Poisson statistical error, as a function of energy and 
for different integration times. In Fig.~\ref{Fig5} (upper panel) are shown 
the values of $R$ obtained from the same set of data (16 pixels, 
$8.2\times10^{4}$~s). From the figure, it is clearly seen that the loss of 
sensitivity per channel does not depend on integration time for 
$\Delta$T~$\ge10$s. 

We have then investigated the effect of these fake events 
in {\em spectral imaging}
data, which is the PICsIT standard mode of operation, using an empty 
field exposure (integration time: 2150s). As the background intensity 
always shows a long-term trend,
expected values of rates were computed by a linear fit to the
data, after selecting an interval of constant linear variation of 
25~hours. For this test, we have used total count rates and verified 
that a large spreading of the values is also present in the energy 
channels below $\approx250$~keV (see Fig.~\ref{Fig5}, lower panel). 

The ratio $R$ changes significantly when rebinning in energy. For example,
we estimate $R\approx4.5$ and $R\approx1.7$ for the energy bands
170-220~keV, and 220-280~keV respectively. These values are sensibly
higher than the ones plotted in Fig.~\ref{Fig5} for the single energy
channels. This is actually expected, as
each track produced by a particle leaves a number of phosphorescent 
counts which are widely distributed in energy, thus giving rise to 
highly correlated count rate fluctuations in different energy channels.

\begin{figure}
\centering
\includegraphics[width=7.5cm,height=13.52cm]{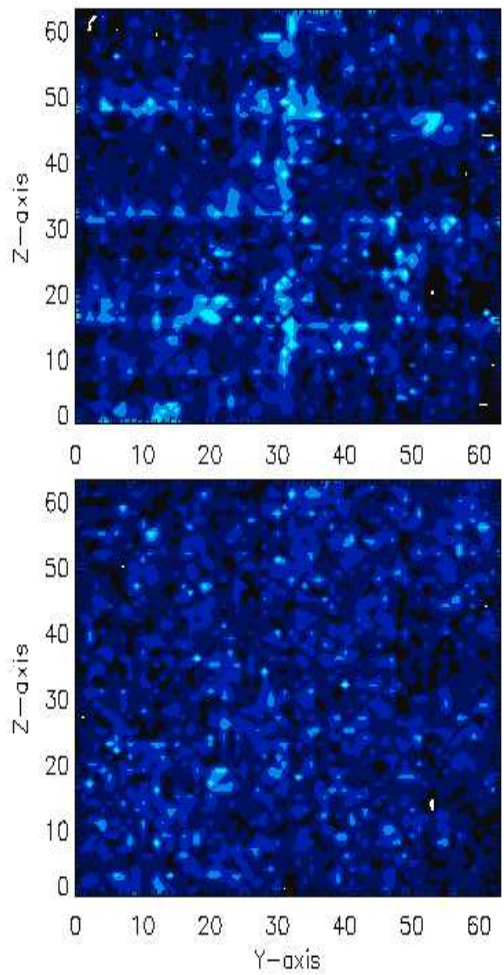}
\caption{The spatial distribution of normal background events in
         PICsIT (upper image), compared to that of phosphorescence triggers in
	 the energy band 170-285 keV (lower image). 
	 The image axes are the same specified in Fig.~\ref{Fig1}.
        }
\label{Fig6}
\end{figure}

Finally, we investigated the spatial distribution of the 
fake events by using the {\em photon by photon} data. 
The images obtained in the energy band 170-285 keV clearly show 
that the non-uniformity in this range cannot be ascribed to the  
phosphorescence triggers (see Fig.~\ref{Fig6}).

\section{Conclusions}
Several effects which reduce the sensitivity of the imaging methods used 
by IBIS, especially in
the PICsIT high energy layer have been described. Some of them are ascribed to
image non-uniformity, and can be corrected by a large extent by a suitable 
flat-field. Some background components (namely, the delayed phosphorescence
emission of the CsI crystals) are found to compromise the sensitivity in the
energy range 180-250 keV, when IBIS is operated in its Standard Mode. 
Image non-uniformity problems are described in the framework of finding 
good methods for background subtraction. The presence of local systematic 
effects due to variable or unexpected behaviour of single pixels and/or ASICs
is hampering the application of {\em standard} flat-field techniques. Present
efforts are devoted to understanding/modelling this type of 
systematics, at the same time refining the already available methods of background 
correction.


\begin{acknowledgements}
The IBIS project is partially granted by Italian Space Agency (ASI). 
AJB is funded by PPARC grant GR/2002/00446.

\end{acknowledgements}

\end{document}